\documentclass[twocolumn,nofootinbib,amsmath]{revtex4}
\usepackage{bbm} 
\usepackage{eufrak} 
\usepackage{mathrsfs} 
\usepackage{tcolorbox}
\usepackage{ragged2e}
\usepackage{array,hhline,multirow}
\usepackage{floatrow,subfig,amssymb}
\usepackage[export]{adjustbox}
\usepackage{rotating} 

\allowdisplaybreaks[2]  
\usepackage{pifont,dsfont}
\usepackage{bm} 

\usepackage{graphicx,tikz}
\usepackage[framemethod=TikZ]{mdframed} 
\usepackage{tikz-cd} 
\usetikzlibrary{arrows,snakes,shapes.arrows,decorations.markings}
     \tikzset{>=triangle 90}
     \tikzstyle{bbc}=[draw,circle,fill=black,scale=.75]
     \tikzstyle{rc}=[circle,fill=red,scale=.6]
     \tikzstyle{wc}=[draw,circle,scale=.75]

\usepackage[colorlinks]{hyperref}
\hypersetup{
    colorlinks = true,
    linkcolor=blue,
    citecolor=magenta,}

\def\bar{\overline}

\def\hat{\widehat}

\def\^{\wedge}

\def\a{{\alpha}}

\def\b{{\beta}}

\def\G{{\Gamma}}

\def\D{{\Delta}}

\def\bh{{\boldsymbol h}}

\def\af{\mathfrak{a}}

\def\cf{\mathfrak{c}}

\def\ff{\mathfrak{f}}
\def\gf{\mathfrak{g}}

\def\Sf{\mathfrak{S}}

\def\spf{\mathfrak{sp}}
\def\suf{\mathfrak{su}}

\def\cD{{\mathcal D}}

\def\cI{{\mathcal I}}

\def\cL{{\mathcal L}}
\def\cM{{\mathcal M}}

\def\cN{{\mathcal N}}
\def\cO{{\mathcal O}}

\def\cS{{\mathcal S}}

\def\cT{{\mathcal T}}

\def\Mm{\cM_{M}[\cT]}

\def\Csc{\mathscr{C}}

\def\H{\mathbb{H}}
\def\N{\mathbb{N}}

\def\Z{\mathbb{Z}} 

\def\beq{\begin{equation}}
\def\eeq{\end{equation}}

\newcommand{\bpmat}{\begin{pmatrix}}
\newcommand{\epmat}{\end{pmatrix}}
\newcommand{\bsmat}{\begin{smallmatrix}}
\newcommand{\esmat}{\end{smallmatrix}}
\newfloatcommand{capbtabbox}{table}[][\FBwidth]

\def\bar{\overline}

\def\hat{\widehat}

\def\^{\wedge}

\def\Mm{\cM_{M}[\cT]}

\def\b{{\beta}}

\def\G{{\Gamma}}

\def\D{{\Delta}}

\def\bh{{\boldsymbol h}}

\def\af{\mathfrak{a}}

\def\cf{\mathfrak{c}}

\def\ff{\mathfrak{f}}
\def\gf{\mathfrak{g}}

\def\Sf{\mathfrak{S}}

\def\spf{\mathfrak{sp}}
\def\suf{\mathfrak{su}}

\def\cD{{\mathcal D}}

\def\cI{{\mathcal I}}

\def\cL{{\mathcal L}}
\def\cM{{\mathcal M}}

\def\cN{{\mathcal N}}
\def\cO{{\mathcal O}}

\def\cS{{\mathcal S}}

\def\cT{{\mathcal T}}

\def\Mm

\def\Csc{\mathscr{C}}

\def\H{\mathbb{H}}
\def\N{\mathbb{N}}

\def\Z{\mathbb{Z}} 

\def\beq{\begin{equation}}
\def\eeq{\end{equation}}

\usepackage{xcolor}

\def\bM{\begin{matrix}}
\def\eM{\end{matrix}}
\def\bar{\overline}

\def\hat{\widehat}

\def\^{\wedge}

\def\fk{\mathfrak{f}}

\def\H{\mathbbm{H}}

\def\cL{{\mathcal L}}
\def\cM{{\mathcal M}}

\def\N{\mathbbm{N}} 
\def\cN{{\mathcal N}}
\def\cO{{\mathcal O}}

\def\cS{{\mathcal S}}

\def\cT{{\mathcal T}}

\def\Z{\mathbbm{Z}}
\def\a{{\alpha}}

\def\b{{\beta}}

\def\G{{\Gamma}}

\def\D{{\Delta}}

\begin{document}
\title{A new rank-2 Argyres-Douglas theory}
\author{Justin Kaidi$^1$}
\author{Mario Martone$^{1,2}$}
\affiliation{1. Simons Center for Geometry and Physics, Stony Brook University}
\affiliation{2. C.~N.~Yang Institute for Theoretical Physics,  Stony Brook University, Stony Brook, NY 11794-3840, USA}


\begin{abstract}
We provide evidence for the existence of a new strongly-coupled four dimensional $\cN=2$ superconformal field theory arising as a non-trivial IR fixed point on the Coulomb branch of the mass-deformed superconformal Lagrangian theory with gauge group $G_2$ and four fundamental hypermultiplets. Notably, our analysis proceeds by using various geometric constraints to bootstrap the data of the theory, and makes no explicit reference to the Seiberg-Witten curve. We conjecture a corresponding VOA and check that the vacuum character satisfies a linear modular differential equation of fourth order. We also propose an identification with existing class $\cS$ constructions. 
\end{abstract}

\maketitle 

\section{Introduction}
Among 4d $\cN=2$ superconformal field theories (SCFTs), of particular interest are those with some chiral ring generators having fractional scaling dimension. Such theories are necessarily non-Lagrangian and as such are not amenable to study by the most naive means. The original examples of such theories, due to Argyres and Douglas \cite{Argyres:1995jj}, can be obtained by tuning to special points in the moduli space of mass-deformed Lagrangian theories, where mutually non-local dyons become massless. Since the early works, several other means of obtaining such theories have been developed, and in particular constructions in class $\cS$ and geometric engineering \cite{Xie:2012hs,Xie:2015rpa,Wang:2018gvb} have allowed one to circumvent the Lagrangian starting point altogether, giving rise to a wealth of Argyres-Douglas (AD) type theories which are not connected in any obvious way to the moduli space of known Lagrangian theories. 

In this note, we proceed in the original spirit of Argyres and Douglas. Namely, we will identify a non-Lagrangian theory obtainable from a mass-deformation of the Lagrangian $G_2$ gauge theory with four fundamental hypermultiplets. However, unlike in the original works, our analysis will make no explicit use of the Seiberg-Witten curve. Instead, we will use a variety of geometric constraints on the structure of the Coulomb and Higgs branches to bootstrap the data of the theory. This will allow us to construct a consistent candidate moduli space, as well as a corresponding VOA. We will also make connection to known class $\cS$ constructions, identifying the resulting theory with the $A_{4}$ 6d (2,0) theory compactified on a sphere with irregular puncture and outer-automorphism twist.  Our example serves as a proof of principle that a bottom-up, geometric approach to bootstrapping general $\cN=2$ SCFTs in 4d is feasible.

\section{Geometric Analysis: Higgs and Coulomb branch}
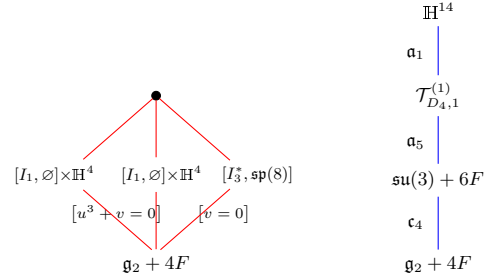
\begin{figure}[tbp]
\begin{center}
\begin{tikzpicture}[decoration={markings,
mark=at position .5 with {\arrow{>}}}]
\begin{scope}[scale=1.5]
\node[bbc,scale=.5] (p0a) at (0,0) {};
\node[scale=.5] (p0b) at (0,-1.4) {};
\node[scale=.8] (t0b) at (0,-1.5){$\mathfrak{g}_2 + 4 F$};
\node[scale=.7] (p1) at (-.8,-.7) {$[I_1,\varnothing]{\times}\H^4$\quad\ \ };
\node[scale=.7] (p2) at (.8,-.7) {\quad\ $[I^*_3,\spf(8)]$};
\node[scale=.7] (p3) at (0,-.6){};
\node[scale=.7] (p4) at (0,-.8){};
\node[scale=.7]  at (.15,-.7) {$[I_1,\varnothing]{\times}\H^4$\quad\ \ };
\node[scale=.8] (t2b) at (-.35,-1.05) {{\scriptsize$\big[u^3+v=0\big]$}};
\node[scale=.8] (t3b) at (.6,-1.05) {{\scriptsize$\big[v=0\big]$}};
\draw[red] (p0a) -- (p1);
\draw[red] (p0a) -- (p2);
\draw[red] (p0a) -- (p3);
\draw[red] (p1) -- (p0b);
\draw[red] (p2) -- (p0b);
\draw[red] (p4) -- (p0b);
\end{scope}
\begin{scope}[scale=1.5,xshift=2.5cm]
\node[scale=.8] (p0a) at (0,.75) {$\H^{14}$};
\node[scale=.8] (t1a) at (-0.2,.37) {$\af_1$};
\node[scale=.8] (p1) at (0,0) {$\cT^{(1)}_{D_4,1}$};
\node[scale=.8] (t1b) at (-.2,-.42) {$\af_5$};
\node[scale=.8] (p2) at (0,-.75) {$\mathfrak{su}(3) + 6F$};
\node[scale=.8] (t1b) at (-.2,-1.13) {$\cf_4$};
\node[scale=.8] (p0b) at (0,-1.5) {$\mathfrak{g}_2 + 4 F$};
\draw[blue] (p0a) -- (p1);
\draw[blue] (p1) -- (p2);
\draw[blue] (p2) -- (p0b);
\end{scope}
\end{tikzpicture}
\caption{Hasse diagrams for the {CB} and {HB} of $\mathfrak{g}_2 + 4 F$.}
\label{G247}
\end{center}
\end{figure}

Our starting point is a Lagrangian SCFT with $G_2$ gauge group and four hypermultiplets in the fundamental ($\mathbf{7}$) representation. The moduli space data of this theory can be conveniently summarized in the Hasse diagrams of Figure \ref{G247}. Here $\H^{n}$ represents the theory of $n$ free hypermultiplets, and the red (resp. blue) lines represent series of singular strata in the Coulomb (resp. Higgs) branches \cite{Martone:2021ixp}. Each Higgs stratum is the minimal nilpotent orbit of a Lie algebra $\gf$, and is denoted by the same symbol.
Following the convention of \cite{Giacomelli:2020jel}, the theory $\cT^{(1)}_{G,1}$ denotes the rank-1 theory obtained on the worldvolume of a D$3$-brane probing a $G$-type exceptional 7-brane. In particular, $\cT^{(1)}_{D_4,1}$ is a $\suf(2)$ gauge theory with $N_f=4$ flavors in the fundamental, while $\cT^{(1)}_{A_2,1}$ and  $\cT^{(1)}_{A_1,1}$, to be introduced momentarily, are the rank-1 AD theories with flavor symmetry $\suf(3)$ and $\suf(2)$, also known as $H_2$ and $H_1$.

The AD theory which we will describe here is a mass-deformation of the $G_2$ gauge theory by 
\begin{align}
\delta \cL = M^{ij} Q_i^\a \Sigma_{\a\b} Q_j^\b~,
\end{align}
where $Q_i^\a$ is a hypermultiplet with flavor index $i=1, \dots, 4$ and fundamental index $\a = 1, \dots, 7$. The matrix $\Sigma_{\a\b}$ is taken to project onto the trivial representation in $\mathbf{7} \otimes \mathbf{7}$ and $ M^{ij}$ is a mass matrix which breaks $\mathfrak{sp}(8)$ to a subgroup $\mathfrak{sp}(4)$.\footnote{\label{footnote1}One could imagine a deformation to a rank-3 flavor symmetry group $\mathfrak{sp}(6)$, but we have not found any theories of this type.} 
Rather than directly studying the effect of this mass deformation, we will proceed by constructing the result from the bottom-up using a variety of geometric constraints.

\begin{figure}
\ffigbox[7cm][]{
\begin{tikzpicture}[decoration={markings,
mark=at position .5 with {\arrow{>}}}]
\begin{scope}[scale=1.5]
\node[bbc,scale=.5] (p0a) at (0,0) {};
\node[scale=.5] (p0b) at (0,-1.4) {};
\node[scale=.8] (t0b) at (0,-1.5) {$\rm AD({{\mathfrak{c}}_2})$};
\node[scale=.7] (p1) at (-.6,-.7) {$[I_1,\varnothing]{\times}\H^2$\quad\ \ };
\node[scale=.7] (p2) at (.6,-.7) {\quad\ $[I^*_1,\spf(4)]$};
\node[scale=.8] (t2b) at (-.5,-1.05) {{\scriptsize$\big[u^5+v^2=0\big]$}};
\node[scale=.8] (t3b) at (.5,-1.05) {{\scriptsize$\big[v=0\big]$}};
\draw[red] (p0a) -- (p1);
\draw[red] (p0a) -- (p2);
\draw[red] (p1) -- (p0b);
\draw[red] (p2) -- (p0b);
\end{scope}
\begin{scope}[scale=1.5,xshift=2.5cm]
\node[scale=.8] (p0a) at (0,.75) {$\H^{4}$};
\node[scale=.8] (t1a) at (-0.2,.37) {$\af_1$};
\node[scale=.8] (p1) at (0,0) {$\cT^{(1)}_{A_1,1}$};
\node[scale=.8] (t1b) at (-.2,-.42) {$\af_1$};
\node[scale=.8] (p2) at (0,-.75) {$(A_1,D_6)$};
\node[scale=.8] (t1b) at (-.2,-1.13) {$\cf_2$};
\node[scale=.8] (p0b) at (0,-1.5) {$\rm AD({{\mathfrak{c}}_2})$};
\draw[blue] (p0a) -- (p1);
\draw[blue] (p1) -- (p2);
\draw[blue] (p2) -- (p0b);
\end{scope}
\end{tikzpicture}}
{\caption{\label{CBADc2} Hasse diagrams for the moduli space of $\rm AD({{\mathfrak{c}}_2})$.}}
\end{figure}
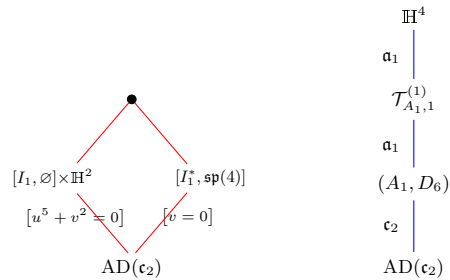

Given two SCFTs $\cT_0$ and $\cT_1$ which are connected by a mass deformation, it is expected that the moduli space of $\cT_1$ has  the same structure of stratification as $\cT_0$, but with the theory on each leaf being replaced with a mass-deformation thereof \cite{Martone:2021drm}.\footnote{There is a subtle but important difference between a symplectic leaf and a symplectic stratum, the latter being the closure of the former.} We recall here the following sequences of mass deformations \cite{Eguchi:1996vu,Argyres:2015ffa}, 
\begin{align}
\mathfrak{su}(3)+6F& \rightarrow D_2(SU(5)) \rightarrow (A_1,D_6) \rightarrow \dots
\nonumber
\\
\cT^{(1)}_{D_4,1} &\rightarrow \cT^{(1)}_{A_2,1} \rightarrow \cT^{(1)}_{A_1,1}\rightarrow \dots
\end{align}
where the series of theories dubbed $D_2(SU(2N+1))$ were studied in \cite{Cecotti:2012jx,Cecotti:2013lda,Xie:2016evu}, while the generalized AD theories $(G_1,G_2)$ were introduced in \cite{Cecotti:2010fi}. This motivates us to propose the Higgs branch Hasse diagram shown in Figure \ref{CBADc2} for our mass-deformed theory. Note that as per footnote \ref{footnote1} we have gone two steps down the sequence of mass deformations. Because we have assumed that the mass-deformation breaks the $\mathfrak{sp}(8)$ flavor symmetry of the starting theory to $\mathfrak{sp}(4)$, we refer to the theory as $\rm AD({{\mathfrak{c}}_2})$.\footnote{We are using the convention in which $\spf(2n)$ is rank $n$. Note also that while AD theories have been historically labeled by the gauge group of the theory on whose Coulomb branch they arise, we are not using this convention. Here ${{\mathfrak{c}}_2}$ refers to the flavor symmetry.} The $(A_1,D_6)$ theory is taken to be supported on the minimal nilpotent orbit of this $\mathfrak{sp}(4)$. That this is the correct thing to do will be confirmed shortly.

\begin{figure}
\capbtabbox[7cm]{%
  \renewcommand{\arraystretch}{1.1}
  \begin{tabular}{|c|c|} 
  \hline
  \multicolumn{2}{|c|}{$\rm AD({{\mathfrak{c}}_2})$}\\
  \hline\hline
  $(\D_u,\D_v)$  &\ ($\frac43,\frac{10}3$)\, \\
  $24a$ &  48 \\  
  $12c$ & 26 \\
$\ff_k$ & $\spf(4)_{\frac{13}3}$ \\ 
$d_{\rm HB}$&4\\
$h$&2\\
$T({\bf2}\bh)$&1\\
\hline\hline
  \end{tabular}
}{%
  \caption{\label{CcADc2} Relevant conformal data for $\rm AD({{\mathfrak{c}}_2})$.}%
}
\end{figure}

Let us denote the unbroken symmetry on a generic point of the nilpotent orbit of $\mathfrak{f}$ by $\fk^\natural$, and the symmetry (level) realized by the theory on the corresponding leaf by $\fk_{\mathrm{IR}} \,\,(k_{\mathrm{IR}})$. Furthermore, say that upon spontaneously breaking $\mathfrak{f}$ to $\fk^\natural$ one produces Goldstone bosons in a representation $\mathfrak{R}$. In terms of this data, the flavor level of the original theory with symmetry $\mathfrak{f}$ is given by \cite{Beem:2019tfp,Beem:2019snk,Giacomelli:2020jel}
\begin{align}
k_{\mathfrak{f}} = {T_2(\mathfrak{R}) +  k_{\mathrm{IR}} I_{\fk_{\mathrm{IR}}\hookrightarrow\fk}  \over I_{{\fk^\natural}\hookrightarrow\fk}}
\end{align}
where $T_2(\mathfrak{R})$ is the Dynkin index of the representation $\mathfrak{R}$ and $I_{\mathfrak{g}\hookrightarrow \fk}$ is the embedding index of $\mathfrak{g}$ into $\fk$. 

In the case of interest to us, $\fk = \mathfrak{sp}(4)$ while $\fk^\natural = \fk_{\mathrm{IR}}= \mathfrak{sp}(2)\cong \suf(2)$. The Goldstone bosons appearing upon spontaneous breaking of $\mathfrak{sp}(4)$ to $\mathfrak{sp}(2)$ are in the $\mathbf{2}$. A standard mathematical exercise then gives $I_{\mathfrak{sp}(2)\hookrightarrow\mathfrak{sp}(4)}=1$ and we choose a normalization for the Dynkin index such that $T_2(\mathbf{2})=1$. Combined with the known flavor level of $(A_1, D_6)$, i.e. $k_{\mathfrak{sp}(2)} = {10 \over 3}$, we conclude that the tentative new theory has 
\begin{align}
k_{\mathfrak{sp}(4)} = {13 \over 3}~.
\end{align}

The structure of the Higgs branch may also be used to constrain the central charges of the new theory. In particular, consider a spontaneous breaking of a flavor symmetry $\ff$ which takes a theory $\cT_0$ to a theory $\cT_{\rm Higgs}$ supported on a leaf $\mathfrak{S}$. Using anomaly matching, the central charges of the two theories can be related by \cite{Giacomelli:2020jel,CCLMW2021}
\begin{align}
12\, c_{\cT_0} = 12\, c_{\cT_{\rm Higgs}}+ 2 \left({3 \over 2}k_\fk - 1 \right) + \mathrm{dim}_{\H}\bar{\mathfrak{S}} - 1~.
\end{align} 
where the bar indicates the closure of $\Sf$ ($\bar{\Sf}$ is then a symplectic stratum). For our purposes, we need only note that the closure of the minimal nilpotent orbit of $\mathfrak{sp}(4)$ has quaternionic dimension $\mathrm{dim}_{\H}\bar{\mathfrak{S}} = 2$. This together with $k_{\mathfrak{sp}(4)}$ computed above, as well as with the central charge $12c_{(A_1, D_6)} = 14$ for the $(A_1, D_6)$ theory, allows us to compute 
\begin{align}
12\, c_{\rm AD({{\mathfrak{c}}_2})} = 26~.
\end{align}

Since the dimension of the stratum $\mathrm{dim}_{\H}\bar{\mathfrak{S}}$ is the difference of the quaternionic dimensions of the parent and daughter Higgs branches, i.e. $\mathrm{dim}_{\H}\bar{\mathfrak{S}} = d_{\mathrm{HB}}(\cT_0) -d_{\mathrm{HB}}(\cT_{\rm Higgs})$, and using the fact that $d_{HB}=2$ for $(A_1, D_6)$, we conclude that the Higgs branch dimension of $\rm AD({{\mathfrak{c}}_2})$ must be 
\begin{align}
d_{\mathrm{HB}}= 4~.
\end{align}

Finally, for theories which flow on the generic point of the Higgs branch to $d_{\mathrm{HB}}$ free hypers, the $c$ and $a$ anomalies are constrained to satisfy
\begin{align}
24(c-a) = d_{\mathrm{HB}}~,
\end{align}
from which we compute 
\begin{align}
24\, a_{\rm AD({{\mathfrak{c}}_2})} = 48~.
\end{align}
The data which we have obtained so far is summarized in Table \ref{CcADc2}.  

Thus far, everything has followed completely from the proposed structure of the Higgs branch. We next turn towards an analysis of the Coulomb branch. The Coulomb branch of a rank-2 theory is a complex two-dimensional space with coordinates $u$ and $v$. These coordinates have scaling dimensions $\Delta_u$ and $\Delta_v$ under a certain $\mathbb{C}^*$-action descending from spontaneously broken dilatations and $\mathfrak{u}(1)_r$. The Coulomb branch again has various strata, which are complex codimension-1 varieties specified by the vanishing loci of polynomials in $u,v$ homogeneous under the $\mathbb{C}^*$-action. Such homogenous polynomials come in two qualitative forms, ``unknotted"  ($u=0,v=0$) and ``knotted" ($u^p + \beta v^q=0$ for $\Delta_v/\Delta_u = p/q$ and $(p,q)=1$) \cite{Argyres:2018zay}. 

Following the notation in \cite{Martone:2020nsy}, we will denote the degree of homogeneity of the polynomial corresponding to the $i$-th stratum $\cI_i$ by $\Delta^{\mathrm{sing}}_i$. This is $\Delta_u$ or $\Delta_v$ for unknotted strata and $p\Delta_u$ for knotted strata. The stratum $\cI_i$ hosts a rank-1 theory, and we will denote the dimension of this theory's Coulumb branch coordinate by $\Delta_i$. Likewise, the central charge, level, and extended Coulomb branch dimension of the theory will be denoted by $c_i,k_i, h_i$.

Let us denote the set of all Coulomb branch strata by $\cI = \left\{\cI_i \right\}$, and the set of strata carrying the UV flavor symmetry by $\cI_{\fk} \subset \cI$. It is a remarkable fact that the central charges and level of the theory at the origin of the Coulomb branch are determined entirely by the data of these strata \cite{Martone:2020nsy}
\begin{align}
\label{eq:Coulombck}
12\, c &\, =\, 4 + h + \sum_{\cI_i\in \cI}{\Delta_i^{\mathrm{sing}} \over \Delta_i} (12\, c_i - h_i - 2)~,
\nonumber\\
k_{\fk} &\, = \,\sum_{\cI_i \in \cI_{\fk}} {\Delta_i^{\mathrm{sing}} \over \Delta_i} [k_i - T(2 h_i)] + T(2h)~,
\end{align} 
together with the relation of Shapere and Tachikawa \cite{Shapere:2008zf}, 
\begin{align}
\label{eq:ShapereTachikawa}
4 (2a-c) = 2 (\Delta_u + \Delta_v) - 2~.
\end{align}
Conversely, given $(c,a,\fk, k_\fk)$ of the theory at the origin, these equations give constraints on the allowed Coulomb branch geometries. 

In addition, there are three more Coulomb branch constraints which must be implemented \cite{Argyres:2018urp,Argyres:2020wmq}. First, the Coulomb branch of any rank-2 theory (excluding products of rank-1 theories), must have at least one knotted stratum. Second, an unknotted stratum $u=0$ (resp. $v=0$) can be present only if $\Delta_v$ (resp. $\Delta_u$) is an allowed rank-1 Coulomb branch dimension, \
\begin{align}
\label{eq:listofrank1dim}
\Delta_{\mathrm{rank-}1}\in\left\{ 6,\, 4,\, 3,\, 2,\, {3\over 2}, \,{4\over 3},\,{6\over 5}\right\}~.
\end{align} 
Finally, according to the UV-IR flavor condition \cite{Martone:2020nsy}, the full flavor symmetry $\fk$ of the theory at the origin must be realized on at least one stratum, i.e. $\cI_\fk$ is non-empty. 

Together, these provide an extremely strict set of constraints, which for generic $(c,a,\fk, k_\fk)$ do not admit a solution. One can interpret our ability to construct a solution at all as non-trivial evidence for our proposal. 

We begin our construction of the Coulomb branch with the UV-IR flavor condition. This requires that at least one of the rank-1 theories describing the effective low-energy theory on the codimension-1 singular loci has to carry flavor symmetry $\mathfrak{sp}(4)$. There exist only two theories with this property \cite{Argyres:2015ffa,Argyres:2015gha,Argyres:2016xua,Argyres:2016xmc,Argyres:2016yzz}, namely $[I_2,\mathfrak{su}(4)]_{\mathbb{Z}_2}$ and $[I_1^*,\mathfrak{sp}(4)]$.\footnote{The former is a $\mathbb{Z}_2$ gauging of $\mathfrak{u}(1)$ gauge theory with $4$ hypermultiplets, while the latter is $\mathfrak{su}(2)$ gauge theory with two hypermultiplets in the $\mathbf{3}$. The latter has a Witten anomaly for the flavor symmetry, while the former does not.} Only the latter can be obtained by deformation of the $[I_3^*,\mathfrak{sp}(8)]$ stratum in Figure \ref{G247}.  Hence this is the case we must choose. We make the minimal guess that there is only a single stratum which carries the $\cf_2$ flavor symmetry, and that this stratum is unknotted, say $v = 0$; this is indeed the case for all known rank-2 theories with simple flavor symmetry \cite{Martone:2021ixp}.  With this, (\ref{eq:ShapereTachikawa}) and the second equation of (\ref{eq:Coulombck}) can be used to determine the dimensions of the Coulomb branch parameters, 
\begin{align}
\Delta_u = {4 \over 3}~, \hspace{0.5 in} \Delta_v = {10 \over 3}~.
\end{align}
Here we have used the known data of the $[I_1^*,\mathfrak{sp}(4)]$ theory, i.e. $(12 c_i, k_i, T(2h_i))= \left({39 \over 4}, 3, 1\right)$. Note that $\Delta_u$ is indeed in the allowed list (\ref{eq:listofrank1dim}) of rank-1 scaling dimensions, as required by our second Coulomb branch constraint.  

To satisfy the first Coulomb branch constraint, we are required to have at least one knotted stratum, and the minimal assumption is to have exactly one knotted stratum carrying no flavor symmetry. There is in fact a unique such possibility compatible with (\ref{eq:Coulombck}), namely the theory $[I_1,\varnothing]$. This completes the construction of the Coulomb branch stratification.

As a final point, note that the dimension $h$ of the extended Coulomb branch of ${\rm AD({{\mathfrak{c}}_2})}$ is equal to that of the theory supported on the stratum realizing the SCFT's flavor symmetry, thus
\begin{align}
h =2~.
\end{align}
This value matches with that expected from the Higgs branch point of view. Indeed, the dimension of the extended Coulomb branch is equal to the total dimension of the symplectic leaves supporting theories of the same rank as our initial SCFT. In the current case, because $(A_1,D_6)$ is of the same rank as AD($\cf_2$), we expect the extended Coulomb branch to be non-trivial and of dimension equal to the dimension of the leaf where $(A_1,D_6)$ is supported, i.e. 2.

In conclusion, the Hasse diagrams in Figure \ref{CBADc2} give a consistent picture of the moduli space for a tentative new theory with data given in Table \ref{CcADc2}. These diagrams pass all non-trivial consistency checks that we know of.

\section{Algebraic analysis: Schur index and VOA}

Having used geometric techniques to derive the data of our tentative theory, we will now make use of a variety of algebraic techniques to check the consistency of this data. 

\subsection{Schur Index and Higgsing}

The Schur index \cite{Gadde:2011uv,Gadde:2011ik} of our proposed theory is given by
\begin{align}
\label{Schur}
\cI_{\rm AD({{\mathfrak{c}}_2})}&={\rm P.E.}\left[{1 \over 1-q}\sum_{i\in 3\N}\left(\chi^{\cf_2}_{\rm adj}\, q^{i+1}+\chi^{\cf_2}_{\bf 4}\,q^{\frac32+i}\right.\right.
\nonumber\\
\vphantom{.}&\qquad\quad\qquad\qquad\left.\left.-\chi^{\cf_2}_{\bf 4}\,q^{\frac52+i}-\chi^{\cf_2}_{\rm adj}\,q^{3+i}\right)\right]
\end{align}
where ${\rm P.E.}$ is the plethystic exponential ${\rm P.E.}[f(x)]:=\exp(\sum_n\tfrac1n f(x^n))$ and $\chi^{\cf_2}_{R}$ is the character for the representation $R$ of $\spf(4)$. 

A first check of this expression is that it reduces to the known Schur index of $(A_1,D_6)$ upon spontaneous breaking of $\spf(4)\rightarrow \suf(2)$. To see this, consider giving a vev to a moment map operator $\cO_{\a_2}$ associated to the long root $\a_2=(-2,2)$ of $\spf(4)$. The contribution of this operator to the index is $q y^{\a_2} := q y_1^2 y_2^{-2}$. Upon Higgsing we remove this contribution, as well as those from $(\cO_{\a_2})^n$ for all $n \in \mathbb{N}$. This roughly amounts to multiplying by $(\sum_n (q y^{\a_2})^n)^{-1} =(1-q y^{\a_2})$ and setting $q y^{\a_2} =1$. More precisely, if $y_2^*$ is the solution to $q y^{\a_2} =1$, then the index after Higgsing is given by \cite{Gaiotto:2012xa,Beem:2014rza,Nishinaka:2018zwq}
\begin{align}
\cI_{\rm Higgs} = \cI_{\rm vec} \lim_{y_2 \rightarrow y_2^*} \left((1-q y^{\a_2})\, \cI_{\rm AD({{\mathfrak{c}}_2})} \right)
\end{align}
where $\cI_{\rm vec}  = (q;q)_{\infty}^2$ is the index of a free vector multiplet and $(a,q)_{\infty}:=\prod_{i=0}^{\infty}(1-a q^i)$ is the \emph{q}-Pochhammer symbol. Stripping off the contribution from two free hypermultiplets and setting $y_1 = 1$, this gives an index with $q$-expansion
\begin{align}
\cI_{(A_1,D_6)} &= (\sqrt{q},q)_{\infty}^2\, \cI_{\rm Higgs}
\nonumber\\
 &= 1+ 4q+ 4q^{3 \over 2}+14q^2 + 16 q^{5\over 2}+\dots
\end{align}
which agrees precisely with the results of \cite{Buican:2015ina}.

\subsection{VOA}
As is well-known \cite{Beem:2013sza}, the Schur index gives the vacuum character of the corresponding VOA via $\chi_{\mathrm{vac}}(q)= q^{\tfrac {c_{4d}}{2}}\cI_{\rm AD({{\mathfrak{c}}_2})}$. In this discussion we will only consider the unrefined version of the index. Setting the flavor fugacities to 1 in (\ref{Schur}) gives the following $q$-expansion:
\beq
\chi_{\rm vac}(q)=q^{\frac{13}{12}}(1+10\, q +4\,q^{\tfrac 32}+65 \,q^2+40\, q^{\tfrac 52}+\dots)~.
\eeq
This vacuum character satisfies a degree-4 twisted modular differential equation $\cD\chi_{\mathrm{vac}}(q) = 0$, where 
\begin{widetext}
\beq\label{mde}
\cD:=D^{(4)}+\left(\frac14\Theta_{1,1}-\frac{11}{24}\Theta_{0,2}\right)D^{(2)}+\left(\frac{1}{72}\Theta_{0,3}-\frac{7}{144}\Theta_{1,2}\right)D^{(1)}+\left(\frac{13}{2304}\Theta_{0,4}-\frac{13}{576}\Theta_{1,3}+\frac{13}{768}\Theta_{2,2}\right)~.
\eeq
\end{widetext}
Here $D^{(n)}$ are the $n$-th order Serre derivatives and $\Theta_{r,s}$ with $s\geq r$ are a basis of weight $k=r+s$ modular forms of $\G^0(2)\subset SL(2,\Z)$. They can be written in terms of Jacobi theta functions $\theta_2$ and $\theta_3$ as follows,
\begin{align}
\Theta_{r,s}(\tau)&:=\theta_2(\tau)^{4r}\theta_4(\tau)^{4s}+\theta_2(\tau)^{4s}\theta_3(\tau)^{4r}~.
\end{align}
For more details on these definitions, see Appendix A of \cite{Beem:2017ooy} and references therein. 

The minimal degree modular differential equation satisfied by $\chi_{\mathrm{vac}}(q)$ is degree-3 and is non-monic. This means that the corresponding VOA is a three-character theory with non-zero Wronskian index. The relevant data for the characters is found to be $(c_{2d}, h_1,h_2)=\left(-26,-{4 \over 3},-{7\over6} \right)$. 
As can be motivated further from the class $\cS$ construction below, a candidate VOA is an extension of affine $\hat{\mathfrak{sp}(4)}_{-{13 \over 6}}$ by four dimension $3 \over 2$ generators. As a first order check, one sees that the usual formula
\begin{align}
c_{2d} = {k_{2d}\, {\rm dim}G \over k_{2d} + h^\vee} = {10 k_{2d} \over k_{2d}+3} = -26
\end{align}
indeed holds for $k_{2d} = - {1 \over 2} k_{4d} = -{13 \over 6}$.  Since this particular level is \emph{admissible} in the language of \cite{Arakawa:2010ni}, it follows that the associated variety is given by the closure of the principal nilpotent orbit of $\mathfrak{sp}(4)$. 

 The Higgs branch identified in Figure \ref{CBADc2} differs from the principal nilpotent orbit of $\mathfrak{sp}(4)$ only in the last stratum, which is hosted on $\mathfrak{a_1} = \mathbb{C}^2/\mathbb{Z}_2$ instead of the expected $\mathbb{C}^2/\mathbb{Z}_4$ \cite{Arakawa:2010ni}. This difference confirms that the VOA is not simply $\hat{\mathfrak{sp}(4)}_{-{13 \over 6}}$, but rather an extension thereof. Indeed, the minimal 4d interpretation of the four operators corresponding to the dimension-$\tfrac 32$ generators featuring in the extension is as Higgs branch operators of $\suf(2)_R$ charge $\frac 32$. These operators are not expected to be nilpotent, and hence are expected to modify the associated variety of $\hat{\mathfrak{sp}(4)}_{-{13 \over 6}}$.\footnote{We thank Leonardo Rastelli and Christopher Beem for illuminating discussions on this point.} 

\subsection{High-temperature limit}

It is by now well-known that the high temperature limit of the $\cN=1$ superconformal index is determined by $c-a$ \cite{DiPietro:2014bca}, from which the high-temperature behavior of the Schur index follows,
\beq\label{Tempera}
\lim_{i\tau \to0}{\rm log}\,\cI_{\rm Schur}(q)\sim \frac{4\pi i(c-a)}{\tau}+{\rm subleading}~.
\eeq
Note that this formula is really only applicable for non-chiral theories with $c>a$ \cite{Ardehali:2015bla,DiPietro:2016ond}. The former is assumed to always be true in the presence of $\cN=2$ supersymmetry, while the latter is obviously true in the case of AD$(\cf_2)$. As explained in \cite{Beem:2017ooy}, the modular differential equation can be used to compute the high-temperature limit of $\cI_{\rm Schur}$. Denoting by $\a_{\rm min}$ the most negative solution of the indicial equation of the S-transformed modular differential equation, the high-temperature limit is given by
\beq
\lim_{i\tau \to0}{\rm log}\,\cI_{\rm Schur}(q)\sim -\frac{4\pi i}{\tau}\frac{\a_{\rm min}}{2}+{\rm subleading}~.
\eeq
We refer to the original literature \cite{Beem:2017ooy} for details.

The expected high-temperature limit of $\cI_{{\rm AD}(\cf_2)}$ can immediately be calculated using \eqref{Tempera} and the values of $a$ and $c$ in Table \ref{CcADc2} to give
\beq
\lim_{i\tau \to0}{\rm log}\,\cI_{{\rm AD}(\cf_2)}\sim \frac{4\pi i}{\tau}\frac16+{\rm subleading}~,
\eeq
which predicts $\a_{\rm min}=-\tfrac 13$. This can indeed be shown to be the most negative solution of the indicial equation of the S-transform of the modular differential equation \eqref{mde}, providing yet another check of our index.

\section{Class $\cS$ construction}
Finally, we close by noting that our theory is obtainable via the class $\cS$ constructions of \cite{Wang:2018gvb}. In particular, consider the type $A_{4}$ 6d (2,0) theory compactified on a two-punctured sphere with one irregular puncture and one maximal regular puncture. To obtain a non-simply laced flavor group, we introduce a $\mathbb{Z}_2$ outer automorphism twist, with the twist line connecting the two punctures. This engineers a class of 4d $\cN=2$ SCFTs denoted by $C^{\mathrm{anom}}_2$, and we choose the one labelled by $k'=-4$ in the notation of \cite{Wang:2018gvb}. The superscript ``$\mathrm{anom}$" here refers to the fact that the $\spf(4)$ flavor symmetry has a Witten anomaly, as follows from \cite{Tachikawa:2018rgw}. With this choice the central charges, flavor levels, and Higgs branch obtained from the formulas in \cite{Wang:2018gvb} match precisely with those for the theory $\rm AD({{\mathfrak{c}}_2})$ discussed here.

It should however be noted that there is a slight mismatch between the VOA predicted by \cite{Wang:2018gvb} and the one we find. 
The analysis of \cite{Wang:2018gvb} suggests that the VOA is a non-extended $\hat{\mathfrak{sp}(4)}_{-{13 \over 6}}$, whereas we predict an extension of this algebra by four dimension-$3 \over 2$ generators. In fact, a similar discrepancy arises in the $C^{\rm anom}_1$ class, where the theory engineered by the irregular singularity with $k'=-1$ corresponds to the rank-2 $H_0$ theory. Again, the VOA of this theory is at an admissible level and the analysis of \cite{Wang:2018gvb} would predict $\hat{\mathfrak{sp}(2)}_{-{17 \over 10}}$ as its VOA. However, the actual VOA of the rank-2 $H_0$ theory was worked out explicitly in \cite{Beem:2019snk} and found to instead be an extension of $\hat{\mathfrak{sp}(2)}_{-{17 \over 10}}$ by two dimension-$\tfrac 52$ generators.  
We take this mismatch as a suggestion that the correct VOA associated to our AD($\cf_2$) is indeed the extended $\hat{\mathfrak{sp}(4)}_{-{13 \over 6}}$.
\newline
\acknowledgments We would like to thank Christopher Beem, Cyril Closset, Simone Giacomelli, Leonardo Rastelli, Shlomo Razamat, Yifan Wang, Gabi Zafrir, and especially Yuji Tachikawa for helpful discussions and comments on the draft. MM is supported by NSF grants PHY-1151392 and PHY-1620610.

\newpage
\bibliographystyle{JHEP}
\bibliography{NewADT}

\providecommand{\href}[2]{#2}\begingroup\raggedright\begin{thebibliography}{10}

\bibitem{Argyres:1995jj}
P.~C. Argyres and M.~R. Douglas, \emph{{New phenomena in SU(3) supersymmetric
  gauge theory}},
  \href{http://dx.doi.org/10.1016/0550-3213(95)00281-V}{\emph{Nucl. Phys.} {\bf
  B448} (1995) 93--126}, [\href{http://arxiv.org/abs/hep-th/9505062}{{\tt
  hep-th/9505062}}].

\bibitem{Xie:2012hs}
D.~Xie, \emph{{General Argyres-Douglas Theory}},
  \href{http://dx.doi.org/10.1007/JHEP01(2013)100}{\emph{JHEP} {\bf 01} (2013)
  100}, [\href{http://arxiv.org/abs/1204.2270}{{\tt 1204.2270}}].

\bibitem{Xie:2015rpa}
D.~Xie and S.-T. Yau, \emph{{4d N=2 SCFT and singularity theory Part I:
  Classification}},  \href{http://arxiv.org/abs/1510.01324}{{\tt 1510.01324}}.

\bibitem{Wang:2018gvb}
Y.~Wang and D.~Xie, \emph{{Codimension-two defects and Argyres-Douglas theories
  from outer-automorphism twist in 6d $(2,0)$ theories}},
  \href{http://dx.doi.org/10.1103/PhysRevD.100.025001}{\emph{Phys. Rev. D} {\bf
  100} (2019) 025001}, [\href{http://arxiv.org/abs/1805.08839}{{\tt
  1805.08839}}].

\bibitem{Martone:2021ixp}
M.~Martone, \emph{{Testing our understanding of SCFTs: a catalogue of rank-2
  $\mathcal{N}$=2 theories in four dimensions}},
  \href{http://arxiv.org/abs/2102.02443}{{\tt 2102.02443}}.

\bibitem{Giacomelli:2020jel}
S.~Giacomelli, C.~Meneghelli and W.~Peelaers, \emph{{New N=2 superconformal
  field theories from S-folds}},  \href{http://arxiv.org/abs/2007.00647}{{\tt
  2007.00647}}.

\bibitem{Martone:2021drm}
M.~Martone and G.~Zafrir, \emph{{On the compactification of 5d theories to
  4d}},  \href{http://arxiv.org/abs/2106.00686}{{\tt 2106.00686}}.

\bibitem{Eguchi:1996vu}
T.~Eguchi, K.~Hori, K.~Ito and S.-K. Yang, \emph{{Study of N=2 superconformal
  field theories in four-dimensions}},
  \href{http://dx.doi.org/10.1016/0550-3213(96)00188-5}{\emph{Nucl. Phys. B}
  {\bf 471} (1996) 430--444}, [\href{http://arxiv.org/abs/hep-th/9603002}{{\tt
  hep-th/9603002}}].

\bibitem{Argyres:2015ffa}
P.~Argyres, M.~Lotito, Y.~L{\"u} and M.~Martone, \emph{{Geometric constraints
  on the space of $ \mathcal{N} $ = 2 SCFTs. Part I: physical constraints on
  relevant deformations}},
  \href{http://dx.doi.org/10.1007/JHEP02(2018)001}{\emph{JHEP} {\bf 02} (2018)
  001}, [\href{http://arxiv.org/abs/1505.04814}{{\tt 1505.04814}}].

\bibitem{Cecotti:2012jx}
S.~Cecotti and M.~Del~Zotto, \emph{{Infinitely many N=2 SCFT with ADE flavor
  symmetry}}, \href{http://dx.doi.org/10.1007/JHEP01(2013)191}{\emph{JHEP} {\bf
  01} (2013) 191}, [\href{http://arxiv.org/abs/1210.2886}{{\tt 1210.2886}}].

\bibitem{Cecotti:2013lda}
S.~Cecotti, M.~Del~Zotto and S.~Giacomelli, \emph{{More on the N=2
  superconformal systems of type $D_p(G)$}},
  \href{http://dx.doi.org/10.1007/JHEP04(2013)153}{\emph{JHEP} {\bf 04} (2013)
  153}, [\href{http://arxiv.org/abs/1303.3149}{{\tt 1303.3149}}].

\bibitem{Xie:2016evu}
D.~Xie, W.~Yan and S.-T. Yau, \emph{{Chiral algebra of Argyres-Douglas theory
  from M5 brane}},  \href{http://arxiv.org/abs/1604.02155}{{\tt 1604.02155}}.

\bibitem{Cecotti:2010fi}
S.~Cecotti, A.~Neitzke and C.~Vafa, \emph{{R-Twisting and 4d/2d
  Correspondences}},  \href{http://arxiv.org/abs/1006.3435}{{\tt 1006.3435}}.

\bibitem{Beem:2019tfp}
C.~Beem, C.~Meneghelli and L.~Rastelli, \emph{{Free Field Realizations from the
  Higgs Branch}}, \href{http://dx.doi.org/10.1007/JHEP09(2019)058}{\emph{JHEP}
  {\bf 09} (2019) 058}, [\href{http://arxiv.org/abs/1903.07624}{{\tt
  1903.07624}}].

\bibitem{Beem:2019snk}
C.~Beem, C.~Meneghelli, W.~Peelaers and L.~Rastelli, \emph{{VOAs and rank-two
  instanton SCFTs}},
  \href{http://dx.doi.org/10.1007/s00220-020-03746-9}{\emph{Commun. Math.
  Phys.} {\bf 377} (2020) 2553--2578},
  [\href{http://arxiv.org/abs/1907.08629}{{\tt 1907.08629}}].

\bibitem{CCLMW2021}
C.~Beem, M.~Martone, C.~Meneghelli, W.~Peelaers and L.~Rastelli, \emph{{A
  bottom up approach for $\cN=2$ SCFTs: rank-1, to appear}}, .

\bibitem{Argyres:2018zay}
P.~C. Argyres, C.~Long and M.~Martone, \emph{{The Singularity Structure of
  Scale-Invariant Rank-2 Coulomb Branches}},
  \href{http://dx.doi.org/10.1007/JHEP05(2018)086}{\emph{JHEP} {\bf 05} (2018)
  086}, [\href{http://arxiv.org/abs/1801.01122}{{\tt 1801.01122}}].

\bibitem{Martone:2020nsy}
M.~Martone, \emph{{Towards the classification of rank-$r$ $\mathcal{N}=2$
  SCFTs. Part I: twisted partition function and central charge formulae}},
  \href{http://arxiv.org/abs/2006.16255}{{\tt 2006.16255}}.

\bibitem{Shapere:2008zf}
A.~D. Shapere and Y.~Tachikawa, \emph{{Central charges of N=2 superconformal
  field theories in four dimensions}},
  \href{http://dx.doi.org/10.1088/1126-6708/2008/09/109}{\emph{JHEP} {\bf 09}
  (2008) 109}, [\href{http://arxiv.org/abs/0804.1957}{{\tt 0804.1957}}].

\bibitem{Argyres:2018urp}
P.~C. Argyres and M.~Martone, \emph{{Scaling dimensions of Coulomb branch
  operators of 4d N=2 superconformal field theories}},
  \href{http://arxiv.org/abs/1801.06554}{{\tt 1801.06554}}.

\bibitem{Argyres:2020wmq}
P.~C. Argyres and M.~Martone, \emph{{Towards a classification of rank $r$
  $\mathcal{N}=2$ SCFTs Part II: special Kahler stratification of the Coulomb
  branch}},  \href{http://arxiv.org/abs/2007.00012}{{\tt 2007.00012}}.

\bibitem{Argyres:2015gha}
P.~C. Argyres, M.~Lotito, Y.~L{\"u} and M.~Martone, \emph{{Geometric
  constraints on the space of $ \mathcal{N} $ = 2 SCFTs. Part II: construction
  of special Kahler geometries and RG flows}},
  \href{http://dx.doi.org/10.1007/JHEP02(2018)002}{\emph{JHEP} {\bf 02} (2018)
  002}, [\href{http://arxiv.org/abs/1601.00011}{{\tt 1601.00011}}].

\bibitem{Argyres:2016xua}
P.~C. Argyres, M.~Lotito, Y.~L{\"u} and M.~Martone, \emph{{Expanding the
  landscape of $ \mathcal{N} $ = 2 rank 1 SCFTs}},
  \href{http://dx.doi.org/10.1007/JHEP05(2016)088}{\emph{JHEP} {\bf 05} (2016)
  088}, [\href{http://arxiv.org/abs/1602.02764}{{\tt 1602.02764}}].

\bibitem{Argyres:2016xmc}
P.~Argyres, M.~Lotito, Y.~L{\"u} and M.~Martone, \emph{{Geometric constraints
  on the space of $ \mathcal{N}$ = 2 SCFTs. Part III: enhanced Coulomb branches
  and central charges}},
  \href{http://dx.doi.org/10.1007/JHEP02(2018)003}{\emph{JHEP} {\bf 02} (2018)
  003}, [\href{http://arxiv.org/abs/1609.04404}{{\tt 1609.04404}}].

\bibitem{Argyres:2016yzz}
P.~C. Argyres and M.~Martone, \emph{{4d $ \mathcal{N} $ =2 theories with
  disconnected gauge groups}},
  \href{http://dx.doi.org/10.1007/JHEP03(2017)145}{\emph{JHEP} {\bf 03} (2017)
  145}, [\href{http://arxiv.org/abs/1611.08602}{{\tt 1611.08602}}].

\bibitem{Gadde:2011uv}
A.~Gadde, L.~Rastelli, S.~S. Razamat and W.~Yan, \emph{{Gauge Theories and
  Macdonald Polynomials}},
  \href{http://dx.doi.org/10.1007/s00220-012-1607-8}{\emph{Commun. Math. Phys.}
  {\bf 319} (2013) 147--193}, [\href{http://arxiv.org/abs/1110.3740}{{\tt
  1110.3740}}].

\bibitem{Gadde:2011ik}
A.~Gadde, L.~Rastelli, S.~S. Razamat and W.~Yan, \emph{{The 4d Superconformal
  Index from q-deformed 2d Yang-Mills}},
  \href{http://dx.doi.org/10.1103/PhysRevLett.106.241602}{\emph{Phys. Rev.
  Lett.} {\bf 106} (2011) 241602}, [\href{http://arxiv.org/abs/1104.3850}{{\tt
  1104.3850}}].

\bibitem{Gaiotto:2012xa}
D.~Gaiotto, L.~Rastelli and S.~S. Razamat, \emph{{Bootstrapping the
  superconformal index with surface defects}},
  \href{http://dx.doi.org/10.1007/JHEP01(2013)022}{\emph{JHEP} {\bf 01} (2013)
  022}, [\href{http://arxiv.org/abs/1207.3577}{{\tt 1207.3577}}].

\bibitem{Beem:2014rza}
C.~Beem, W.~Peelaers, L.~Rastelli and B.~C. van Rees, \emph{{Chiral algebras of
  class S}}, \href{http://dx.doi.org/10.1007/JHEP05(2015)020}{\emph{JHEP} {\bf
  05} (2015) 020}, [\href{http://arxiv.org/abs/1408.6522}{{\tt 1408.6522}}].

\bibitem{Nishinaka:2018zwq}
T.~Nishinaka, S.~Sasa and R.-D. Zhu, \emph{{On the Correspondence between
  Surface Operators in Argyres-Douglas Theories and Modules of Chiral
  Algebra}}, \href{http://dx.doi.org/10.1007/JHEP03(2019)091}{\emph{JHEP} {\bf
  03} (2019) 091}, [\href{http://arxiv.org/abs/1811.11772}{{\tt 1811.11772}}].

\bibitem{Buican:2015ina}
M.~Buican and T.~Nishinaka, \emph{{On the superconformal index of
  Argyres\textendash{}Douglas theories}},
  \href{http://dx.doi.org/10.1088/1751-8113/49/1/015401}{\emph{J. Phys. A} {\bf
  49} (2016) 015401}, [\href{http://arxiv.org/abs/1505.05884}{{\tt
  1505.05884}}].

\bibitem{Beem:2013sza}
C.~Beem, M.~Lemos, P.~Liendo, W.~Peelaers, L.~Rastelli and B.~C. van Rees,
  \emph{{Infinite Chiral Symmetry in Four Dimensions}},
  \href{http://dx.doi.org/10.1007/s00220-014-2272-x}{\emph{Commun. Math. Phys.}
  {\bf 336} (2015) 1359--1433}, [\href{http://arxiv.org/abs/1312.5344}{{\tt
  1312.5344}}].

\bibitem{Beem:2017ooy}
C.~Beem and L.~Rastelli, \emph{{Vertex operator algebras, Higgs branches, and
  modular differential equations}},
  \href{http://dx.doi.org/10.1007/JHEP08(2018)114}{\emph{JHEP} {\bf 08} (2018)
  114}, [\href{http://arxiv.org/abs/1707.07679}{{\tt 1707.07679}}].

\bibitem{Arakawa:2010ni}
T.~Arakawa, \emph{{Associated varieties of modules over Kac-Moody algebras and
  C(2)-cofiniteness of W-algebras}},
  \href{http://arxiv.org/abs/1004.1554}{{\tt 1004.1554}}.

\bibitem{DiPietro:2014bca}
L.~Di~Pietro and Z.~Komargodski, \emph{{Cardy formulae for SUSY theories in $d
  =$ 4 and $d =$ 6}},
  \href{http://dx.doi.org/10.1007/JHEP12(2014)031}{\emph{JHEP} {\bf 12} (2014)
  031}, [\href{http://arxiv.org/abs/1407.6061}{{\tt 1407.6061}}].

\bibitem{Ardehali:2015bla}
A.~Arabi~Ardehali, \emph{{High-temperature asymptotics of supersymmetric
  partition functions}},
  \href{http://dx.doi.org/10.1007/JHEP07(2016)025}{\emph{JHEP} {\bf 07} (2016)
  025}, [\href{http://arxiv.org/abs/1512.03376}{{\tt 1512.03376}}].

\bibitem{DiPietro:2016ond}
L.~Di~Pietro and M.~Honda, \emph{{Cardy Formula for 4d SUSY Theories and
  Localization}}, \href{http://dx.doi.org/10.1007/JHEP04(2017)055}{\emph{JHEP}
  {\bf 04} (2017) 055}, [\href{http://arxiv.org/abs/1611.00380}{{\tt
  1611.00380}}].

\bibitem{Tachikawa:2018rgw}
Y.~Tachikawa, Y.~Wang and G.~Zafrir, \emph{{Comments on the twisted punctures
  of $A_\text{even}$ class S theory}},
  \href{http://dx.doi.org/10.1007/JHEP06(2018)163}{\emph{JHEP} {\bf 06} (2018)
  163}, [\href{http://arxiv.org/abs/1804.09143}{{\tt 1804.09143}}].

\end{thebibliography}\endgroup
\end{document}